\documentclass[conference]{IEEEtran}
\IEEEoverridecommandlockouts

\usepackage{cite}
\usepackage{amsmath,amssymb,amsfonts}
\usepackage{algorithm}
\usepackage{algorithmic}
\usepackage{graphicx}
\usepackage{textcomp}
\usepackage{xcolor}
\usepackage[hyphens]{url}
\usepackage{booktabs}
\usepackage{multirow}
\usepackage{colortbl}

\def\BibTeX{{\rm B\kern-.05em{\sc i\kern-.025em b}\kern-.08em
    T\kern-.1667em\lower.7ex\hbox{E}\kern-.125emX}}

\definecolor{RobustTeal}{HTML}{2AA7A0}
\definecolor{FragilePurple}{HTML}{7F77DD}

\newcommand{\Intact}{\text{Intact}}
\newcommand{\Tampered}{\text{Tampered}}
\definecolor{glist}{HTML}{CDEAD7}   % green-list token background
\definecolor{rlist}{HTML}{F7D1D1}   % red-list token background
\definecolor{editc}{HTML}{E07A2F}   % adversarial-edit highlight (orange)
\definecolor{framegray}{gray}{0.3} % spoof-example figure border
\newcommand{\gtok}[1]{\colorbox{glist}{\vphantom{Ay}#1}}
\newcommand{\rtok}[1]{\colorbox{rlist}{\vphantom{Ay}#1}}
\newcommand{\egtok}[1]{{\setlength{\fboxrule}{1pt}\fcolorbox{editc}{glist}{\vphantom{Ay}\textbf{#1}}}}
\newcommand{\ertok}[1]{{\setlength{\fboxrule}{1pt}\fcolorbox{editc}{rlist}{\vphantom{Ay}\textbf{#1}}}}

\newcommand{\zr}{z_{r}}
\newcommand{\zf}{z_{f}}
\newcommand{\suf}{\mathrm{suffix}}

\definecolor{editbg}{HTML}{FDEBDD}
\definecolor{homoc}{HTML}{4A7FB5}
\definecolor{homobg}{HTML}{E2ECF6}
\newcommand{\etok}[1]{{\setlength{\fboxrule}{1pt}\fcolorbox{editc}{editbg}{\vphantom{Ay}\textbf{#1}}}}
\newcommand{\htok}[1]{{\setlength{\fboxrule}{1pt}\fcolorbox{homoc}{homobg}{\vphantom{Ay}#1}}}

\title{Tracing Provenance and Detecting Tampering with Complementary LLM Watermarks}

\author{
\IEEEauthorblockN{Xiaoyan Feng, Yanjun Zhang, He Zhang, Leo Yu Zhang, and Shirui Pan}
\IEEEauthorblockA{\textit{Griffith University}}
}

\begin{document}

\maketitle

\begin{abstract}
Watermarking LLM-generated text is an important task for tracing its provenance.
Existing LLM watermarks preserve provenance under editing, but this same robustness allows an adversary to alter critical content while retaining attribution, a vulnerability known as piggyback spoofing. 
We introduce an innovative watermark that jointly provides provenance and tamper evidence. 
It co-embeds a robust signal and a fragile signal into each generated token. 
The signals share the same mechanism but use independent keys and different seeding windows over normalized text, making one resilient to edits and the other sensitive to reader-visible changes. 
Multiple rounds of unbiased tournament reweighting preserve the expected generation distribution, while a periodic round-allocation pattern controls the trade-off between the two signals. 
At detection, their scores form a two-dimensional space supporting three decisions: Intact, Tampered, and No-Watermark. 
Across two large language models and two prompt datasets, our method demonstrates the highest tamper-detection rate among the evaluated methods while maintaining competitive attribution robustness and perplexity.
Ablation studies show that reliable three-state detection requires a well-defined notion of intactness, co-embedding of the two signals, and complementary sensitivity to edits. 
\end{abstract}


\section{Introduction}

\begin{figure*}[t!]
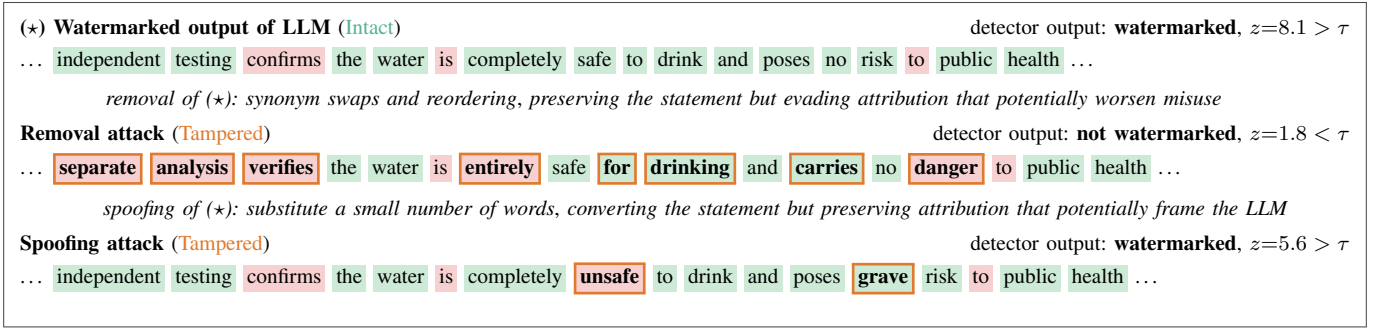

\centering
\footnotesize
{\setlength{\fboxsep}{6pt}%
\fcolorbox{framegray}{white}{%
\begin{minipage}{0.97\textwidth}
\setlength{\fboxsep}{1.4pt}%
\textbf{($\star$) Watermarked output of LLM} (\textcolor[HTML]{59B292}{\Intact})\hfill detector output: \textbf{watermarked}, $z{=}8.1>\tau$\\[3pt]
\ldots\ \gtok{independent} \gtok{testing} \rtok{confirms} \gtok{the} \gtok{water} \rtok{is} \gtok{completely} \gtok{safe} \gtok{to} \gtok{drink} \gtok{and} \gtok{poses} \gtok{no} \gtok{risk} \rtok{to} \gtok{public} \gtok{health}\ \ldots
\par\vspace{6pt}
{\centering \quad \emph{removal of ($\star$): synonym swaps and reordering}, \emph{preserving the statement but evading attribution that potentially worsen misuse\qquad\quad\quad}\par}
\vspace{4pt}
\textbf{Removal attack} (\textcolor[HTML]{E07A2F}{\Tampered{}})\hfill detector output: \textbf{not watermarked}, $z{=}1.8<\tau$\\[3pt]
\ldots\ \ertok{separate} \ertok{analysis} \ertok{verifies} \gtok{the} \gtok{water} \rtok{is} \ertok{entirely} \gtok{safe} \egtok{for} \egtok{drinking} \gtok{and} \egtok{carries} \gtok{no} \ertok{danger} \rtok{to} \gtok{public} \gtok{health}\ \ldots
\par\vspace{6pt}
{\centering \quad \emph{spoofing of ($\star$): substitute a small number of words}, \emph{converting the statement but preserving attribution that potentially frame the LLM}\par}
\vspace{4pt}
\textbf{Spoofing attack} (\textcolor[HTML]{E07A2F}{\Tampered{}})\hfill detector output: \textbf{watermarked}, $z{=}5.6>\tau$\\[3pt]
\ldots\ \gtok{independent} \gtok{testing} \rtok{confirms} \gtok{the} \gtok{water} \rtok{is} \gtok{completely} \ertok{unsafe} \gtok{to} \gtok{drink} \gtok{and} \gtok{poses} \egtok{grave} \gtok{risk} \rtok{to} \gtok{public} \gtok{health}\ \ldots
\par\vspace{6pt}
\end{minipage}}}
\caption{An illustrative example of removal and spoofing attacks on a green--red list watermark~\cite{kgw}, with green-token z-score $z$ and detection threshold $\tau$.  
Token shading marks \gtok{green-list} or \rtok{red-list} membership under the secret key.
}
\label{fig:spoof-example}
\end{figure*}

Regulations and industry commitments increasingly require AI-generated content to be identifiable~\cite{eu_ai_act,ca_ai_act,wh_commitments}.
Generative watermarking is a principled path to this requirement, tracing the \emph{provenance} of LLM-generated text~\cite{kgw,synthid}.
It tilts the sampling probabilities toward tokens favored by a keyed pseudorandom signal, and later scores how strongly the received text follows this signal.
The first threat to watermarking is removal, in which an attacker edits part of the watermarked text yet still evades provenance attribution.
Existing schemes therefore keep the signal robust to edits to ensure reliable provenance~\cite{unigram,sir}.
This robustness opens up a new attack surface.
An adversary may alter critical details of watermarked text while keeping the signal strong, making the scheme endorse content the model did not generate.
This attack is called \emph{piggyback spoofing}~\cite{nofreelunch}.
Figure~\ref{fig:spoof-example} illustrates both attacks on the same green-red watermark~\cite{kgw}.
Countering piggyback spoofing requires \emph{tamper evidence}.
Tamper evidence calls for a signal sensitive to edits, which is the opposite of what provenance demands.
Yet most schemes embed a single signal and thus face a dilemma.

\begin{figure}
  \centering
  \includegraphics[width=1.0\columnwidth]{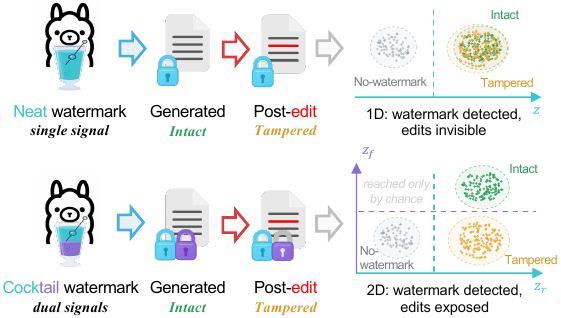}
  \caption{
Cocktail co-embeds two signals into one text: a \textcolor{RobustTeal}{\textbf{robust signal}} and a \textcolor{FragilePurple}{\textbf{fragile signal}}.
A lightly edited text keeps a high robust score but drops to a low fragile score, making tamper evident while preserving provenance.
   }
  \label{fig:teaser}
\end{figure}

Embedding two signals in one text decouples provenance and tamper evidence and promises both.
A watermark scheme fulfilling this goal requires a tailored signal design, embedding mechanism, and detection rule.
The closest attempt, Bileve~\cite{bileve}, pays a price at every stage.
It signs the first tokens of a generation and writes the signature bits into the following tokens.
The carrier tokens are generated after the signature is computed and cannot enjoy its protection.
At detection, its signature fails even on unedited text, because re-tokenization of the delivered text may shift the token IDs.
Generation quality also degrades, with perplexity rising about sixfold, because the sampler takes the top-ranked token whose hash matches the signature bit.

We propose Cocktail, which embeds two complementary signals that differ only in the key and the seeding window, achieving provenance and tamper evidence simultaneously.
The signal for provenance is robust to edits via a short window, while the signal for tamper evidence is fragile to edits via a long window.
The signals are seeded on normalized text to ensure they react to reader-visible edits but stay consistent during transmission.
Both signals are co-embedded into each token, so that every token contributes to both demands.
The co-embedding is implemented by vectorized tournament reweighting, with a periodic pattern controlling the signal strength ratio while maintaining generation quality.
In detection, the two signal scores span a 2D plane where Intact, Tampered, and No-Watermark text occupy separable regions, as Figure~\ref{fig:teaser} previews.
A two-threshold rule then reads the three states directly off the plane.
Our contributions are fourfold.
\begin{itemize}
  \item We resolve the dilemma by decoupling: a robust and a fragile signal in one text serve provenance and tamper evidence separately, and their joint score turns the binary decision into a three-state decision.
  \item We define tamper evidence over the content the reader receives rather than the token IDs that carry it, which lays the foundation for the three-state decision.
  \item We co-embed the two signals into every token through rounds of unbiased reweighting, costing neither generation quality nor detection efficiency, and provide a tunable strength ratio between the signals.
  \item Experiments show that Cocktail leads the strongest single-signal baseline by over 66 percentage points in tamper evidence without conceding attribution, and ablations trace this margin to each design choice above.
\end{itemize}

\section{Related Work}
\subsection{Generative Watermarks for LLM Text}
Aaronson~\cite{aaronson} first involved a keyed pseudorandom variable in token sampling, making the sampled token indices statistically correlated with it.
KGW~\cite{kgw} instantiated the signal as a green-red partition of the vocabulary, boosting green tokens at sampling and scoring text by the z-score of its green-token count.
Together they established the framework of inference-time embedding and detection.
Dipper~\cite{dipper} fine-tuned T5-XXL~\cite{T5} into a paraphraser and showed that embedded watermarks withstand paraphrasing better than passive detectors~\cite{mitchell2023detectgpt}.
Later work pushes two directions: robustness to edits and generation quality.

Toward robustness, Unigram~\cite{unigram} fixes one global partition to maximize edit resilience; Kuditipudi et al.~\cite{robustdf} align text to a pregenerated key sequence by Levenshtein distance; SIR~\cite{sir}, SemStamp~\cite{semstamp}, and further semantics-based schemes~\cite{semantics_paraphrase_wm,adaptive_wm} seed the signal on semantics to survive paraphrasing.
The pursuit has a proven ceiling: strong watermarking against unbounded rewriting is impossible~\cite{sand_impossibility}, and reliability degrades with the edit budget~\cite{kirchenbauer_reliability}.

Toward generation quality, the distortion-free notion, formulated in~\cite{robustdf} and independently in~\cite{undetectability,unbiased}, requires the expected token probability under the randomized signal to equal the original one.
Unbiased~\cite{unbiased} further shows that this preserves expected downstream performance provided the seeds do not repeat.
DiPmark~\cite{dipmark}, SynthID~\cite{synthid}, MCmark~\cite{mcmark}, and BiMark~\cite{bimark} design their own unbiased reweightings for signal embedding.
BiMark and ENS~\cite{ens_unbiased} compose independent unbiased steps to amplify one signal, whereas Cocktail composes tournament rounds, whose accountable entropy consumption turns round allocation into a strength dial between two complementary signals.

\begin{figure*}[t!]
  \centering
  \includegraphics[width=1.0\textwidth]{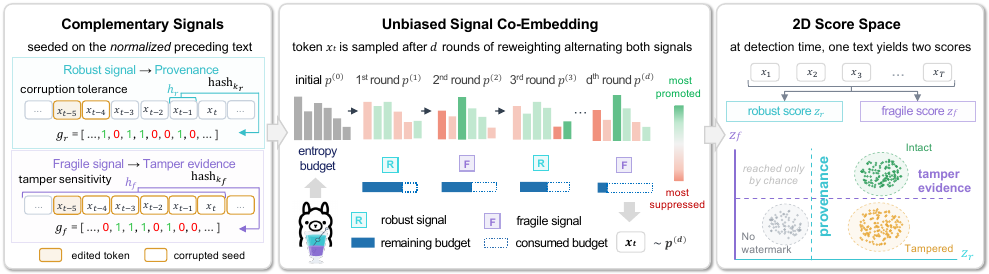}
  \caption{Cocktail Overview.
  \textbf{(1) Complementary Signals.} Both signals are seeded on the normalized preceding text, with a short robust window of $h_r$ tokens and a long fragile window of $n_f$ characters.
  \textbf{(2) Unbiased Signal Co-Embedding.} Each step applies rounds of unbiased reweighting, allocated between the two signals, preserving the expected output distribution.
  \textbf{(3) 2D Score Space.} The two scores $\zr$ and $\zf$ place the three states in distinct regions, classified by a joint rule.
  }
  \label{fig:overview}
\end{figure*}

\subsection{Piggyback Spoofing and Its Defenses}
Piggyback spoofing~\cite{nofreelunch} modifies a small number of words in an existing watermarked text and exploits the robustness of the signal so that the edited text retains a high signal strength, exposing a vulnerability in existing schemes.
Bileve~\cite{bileve}, the work closest to Cocktail, attempts to counter this attack.
Bileve embeds a coarse-grained and a fine-grained signature, but the signature distorts the output to carry its bits and certifies token IDs rather than the text the reader receives.
An et al.~\cite{an2025} tighten the semantic binding of the signal through a second-pass generation, yet it remains a single signal that still faces the conflict between robust attribution and spoofing resistance.
Escaping this conflict calls for complementary signals, a principle realized in image watermarking by pairing a robust and a fragile watermark, under the same name~\cite{lu_cocktail_2000}.
Cocktail, named for mixing the two signals round by round into every token, realizes this principle for generative LLM text.

\section{Problem Setup}
An autoregressive language model generates text token by token. 
At step $t$, it outputs a distribution $p(\cdot\mid x_{<t})$ over the vocabulary $\mathcal{V}$ and samples the next token $x_t$.
A generative watermark adjusts this distribution with a pseudorandom signal derived from a secret key and a seed.
Detection receives only the text, recomputes the signal from the text and the key, and scores how strongly the text follows it.

Between generation and detection, the text passes through hands that may change it.
The adversary knows the scheme but not the keys~\cite{cayre_security}.
It receives watermarked text and edits it, aiming either to remove the watermark while keeping the main content, or to alter critical details without being noticed by detection.
The detector holds the keys but no access to the language model.

Under this threat, detection faces two questions.
\emph{Provenance} asks whether the text originates from the watermarked model.
\emph{Tamper evidence} asks whether the delivered text is exactly what the model generated.

\section{Method}
\subsection{Overview}
Cocktail provides both provenance and tamper evidence, implemented through three mechanisms shown in Figure~\ref{fig:overview}.
The two goals are achieved by two \textbf{complementary signals}.
The robust signal serves provenance, and the fragile signal serves tamper evidence.
The hash input of both signals is the content the watermark is anchored to: the normalized text.
The two signals enter the generation distribution of every token through \textbf{unbiased co-embedding}, an adaptive probability reweighting applied round by round.
Each round of reweighting consumes a share of the collision entropy of the input distribution.
We define a periodic rule for the embedding where the two signals are alternated to distribute the strength between them.
The two signal scores span a \textbf{two-dimensional score plane}, where post-edit watermarked text decays fast along the fragile axis but slowly along the robust axis, and this asymmetry exposes it.
Cocktail thereby supports a three-state decision.

\subsection{Complementary Watermark Signals}

Cocktail embeds a \emph{robust signal} and a \emph{fragile signal} that serve the two complementary goals of provenance and tamper evidence.
We call a signal robust or fragile by its sensitivity to post-edits.
We characterize a watermark signal by its seed and its carrier.
For generative watermarks, the carrier is what is sampled from the signal-adjusted distribution, and the seed is what a hash function takes to generate the signal.

\paragraph{Independent green-red list.}
Concretely, each signal is a green-red list over the vocabulary, imprinted on the sampling probabilities of the tokens.
We denote the robust signal as $g_r$ and the fragile signal as $g_f$.
For a token $x_i\in\mathcal{V}$, $g_r[i]=1$ means that the robust signal assigns this token green, and $g_r[i]=0$ means red.
The fragile signal $g_f$ follows the same form.
Formally, at generation step $t$, each signal is a pseudorandom function of its key and its seed,
\begin{equation}\label{eq:signal}
  g_\ast^{(t)}=\mathrm{PRNG}\bigl(k_\ast,\, s_\ast(t)\bigr)\in\{0,1\}^{|\mathcal{V}|},\quad \ast\in\{r,f\},
\end{equation}
where the key $k_\ast$ identifies the signal and the seed $s_\ast(t)$ is specified below.
Following prior work, the signal raises the generation probability of its green tokens and suppresses that of its red tokens.
The two signals are generated under different keys, $k_r\neq k_f$, and therefore mutually independent.

\paragraph{Short window versus long window.}
Provenance requires a signal that survives edits, and tamper evidence requires a signal that breaks under them.
The window length sets this sensitivity.
An edit at position $i$ corrupts the seed $s_\ast(t)$ at every $t\in[i,\ i+h]$ for a window of length $h$.
A short window confines the damage to a few tokens, and a long window spreads it far downstream.
Existing schemes set the window between one and three tokens and thus sit at the robust end alone.
We therefore assign the robust signal a short window $h_r$ and the fragile signal a long window $n_f$.
The left panel of Figure~\ref{fig:overview} depicts how the two windows respond to an edit.

\paragraph{Seeded on normalized text.}
In most previous work, token IDs serve as both the seed and the carrier.
We keep tokens as the carrier yet take the normalized text as the seed.
The normalized seed makes tamper evidence for watermarked text well-defined.
Token ids are intermediates of generation, and the content delivered to the reader is the normalized text.
Formally, with $\mathcal{N}$ the normalizer and $H$ a hash function,
\begin{equation}\label{eq:seed}
\begin{aligned}
  s_r(t)=H\bigl(\mathcal{N}(x_{\max(1,\,t-h_r):t-1})\bigr),\\
  s_f(t)=H\bigl(\suf_{n_f}\bigl(\mathcal{N}(x_{1:t-1})\bigr)\bigr),
\end{aligned}
\end{equation}
where $\suf_{n}$ keeps the last $n$ characters and both windows truncate at the text start when the prefix is shorter than the window.
Detection re-tokenizes the delivered text, and the drifted token IDs realign within a few tokens for a short window $h_r$.
The operator $\suf_{n_f}$ slides a character window over the growing normalized prefix.
On intact text, the normalized suffix is the same string at generation and at detection, and the fragile signal thus reacts only to deliberate edits.

Specifically, our normalizer includes invisible-character cleanup, Unicode \textsc{nfkc}, homoglyph folding to canonical characters, and whitespace collapsing.
These operations are idempotent and reproducible across environments.
Replacing the normalizer only widens or narrows the range of perturbations within which the seeds stay reproducible, leaving the co-embedding design untouched.
The analogy is a signed document, where the signature certifies the wording rather than the paper it is written on.

\subsection{Unbiased Signal Co-Embedding}

To maximize both provenance and tamper evidence, both robust and fragile signals are expected to be embedded into each token.
For practicality, the co-embedding aims to preserve generation quality, and keep the strength ratio between the two signals tunable.

\paragraph{Multiple round unbiased reweightings}.
A token carries both signals when both signals shape its sampling probability. Cocktail therefore adjusts the distribution using the two signals in sequence, as shown in the middle panel of Figure~\ref{fig:overview}.
Repeated adjustment is where quality is at stake: two adjustments of a biased reweighting drift the distribution further from its expectation.
Cocktail preserves the quality by using an unbiased reweighting by which the expected distribution is maintained even after multiple rounds adjustment.

Cocktail uses vectorized tournament, a closed form of the two-sample tournament sampling of SynthID~\cite{synthid}.
Let $\hat{p}^{(0)} = p$ and let $d$ denote the number of rounds. 
In round $i$, an independent green-red list $g^{(i)}[x] \sim \mathrm{Bernoulli}(0.5)$ reweights the previous round's output:
\begin{equation}\label{eq:vt}
  \begin{aligned}
  &\hat{p}^{(i)}(x) = \bigl(1 + g^{(i)}[x] - q^{(i)}\bigr)\,
  \hat{p}^{(i-1)}(x),
  \qquad \\
  &q^{(i)} = \sum_{x \in \mathcal{V}} g^{(i)}[x]\,\hat{p}^{(i-1)}(x),
  \end{aligned}
\end{equation}
and the token is sampled from the final distribution $\hat{p}^{(d)}$.
Here, $\hat{p}^{(i)}$ is the reweighted distribution, and $q^{(i)}$ is the total probability of the green tokens under $\hat p^{(i-1)}$.
A green token with $g^{(i)}[x]=1$ gains a fraction $1-q^{(i)}$ of its probability, while a red token with $g^{(i)}[x]=0$ loses a fraction $q^{(i)}$.
Since $g^{(i)}[x]\sim\mathrm{Bernoulli}(0.5)$, we have $\mathbb{E}\bigl[g^{(i)}[x]\bigr]=\mathbb{E}\bigl[q^{(i)}\bigr]=\tfrac12$, and each token's expected adjustment factor is $\mathbb{E}\bigl[1+g^{(i)}[x]-q^{(i)}\bigr]=1$, i.e., the reweighting is unbiased.
Because the lists are independent across rounds, unbiasedness is preserved after all $d$ rounds, i.e., $\mathbb{E}\bigl[\hat{p}^{(d)}(x)\bigr]=p(x)$.

\begin{algorithm}[t]
\caption{Cocktail Watermark: Embedding}
\label{alg:embed}
\begin{algorithmic}[1]
\REQUIRE model $\mathsf{M}$, prompt $a$, keys $k_r, k_f$, number of
  rounds $d$, pattern $\pi$, short window $h_r$, fragile character
  window $n_f$, length $T$, normalizer $\mathcal{N}$, hash $H$.
\STATE $x \leftarrow \emptyset$
\FOR{$t = 1, \dots, T$}
  \STATE $\hat{p} \leftarrow \mathsf{M}(\cdot \mid a,\, x)$
  \FOR{$i = 1, \dots, d$}
    \STATE $\ast \leftarrow \pi_i$;\;\;
      $c \leftarrow \mathcal{N}(x_{\max(1,\,t-h_r):t-1})$ \textbf{if}
      $\ast{=}r$ \textbf{else} $\suf_{n_f}(\mathcal{N}(x_{1:t-1}))$
    \STATE $g^{(i)} \leftarrow
      \mathrm{PRNG}\bigl(k_\ast,\, H(c \,\|\, i)\bigr)$
    \STATE $\hat{p} \leftarrow
      \bigl(1 + g^{(i)} - q^{(i)}\bigr)\,\hat{p}$
      \hfill $\triangleright$ Eq.~\eqref{eq:vt}
  \ENDFOR
  \STATE $x_t \sim \hat{p}$;\; \textbf{break if} $x_t$ is a stop
    token;\; $x \leftarrow x \,\|\, x_t$
\ENDFOR
\RETURN $x$
\end{algorithmic}
\end{algorithm}

\paragraph{Periodic pattern and strength ratio.}
Vectorized tournament supports tuning the strength ratio between the two signals by allocating rounds in a periodic pattern.
We assign each round to one signal according to a pattern $\pi = (\pi_1, \dots, \pi_d)$ with $\pi_i \in \{\textsc{Robust}, \textsc{Fragile}\}$.
Given a robust-to-fragile ratio $(r{:}1)$, the pattern is periodic with period $r+1$:
\begin{equation}\label{eq:pattern}
  \pi_i =
  \begin{cases}
    \textsc{Fragile} & i \bmod (r+1) = 0,\\
    \textsc{Robust}  & \text{otherwise},
  \end{cases}
  \qquad i = 1, \dots, d,
\end{equation}
yielding $d_r=|\{i : \pi_i = \textsc{Robust}\}|$ robust rounds and $d_f = d - d_r$ fragile rounds.
In round $i$, the green-red list is drawn under the designated signal's key with the round index folded into the hash, $g^{(i)}=\mathrm{PRNG}\bigl(k_{\pi_i},\, H(c \,\|\, i)\bigr)$, where $c$ is the seeding window content of that signal, and the lists stay independent across rounds even within one signal.
Balanced alternation corresponds to the pattern $(1{:}1)$.

The round ratio approximates the strength ratio because the collision entropy~\cite{cover_elements} of the initial distribution acts as a budget shared by the two signals.
Each round spends a share of the remaining budget, and rounds allocated at $(r{:}1)$ split the budget roughly at that ratio.
The proxy is not exact, since the consumption per round is nonlinear and the strength does not grow linearly with the budget, yet the round ratio remains a monotone and practical knob for adjusting the strength ratio.

\begin{algorithm}[t]
\caption{Cocktail Watermark: Detection}
\label{alg:detect}
\begin{algorithmic}[1]
\REQUIRE text $x = (x_1, \dots, x_T)$, keys $k_r, k_f$, number of
  rounds $d$, pattern $\pi$, thresholds $\tau_r, \tau_f$, short window
  $h_r$, fragile character window $n_f$, normalizer $\mathcal{N}$, hash $H$.
\STATE $N_r, N_f \leftarrow 0$
\FOR{$t = 1, \dots, T$}
  \FOR{$i = 1, \dots, d$}
    \STATE $\ast \leftarrow \pi_i$;\;\;
      $c \leftarrow \mathcal{N}(x_{\max(1,\,t-h_r):t-1})$ \textbf{if}
      $\ast{=}r$ \textbf{else} $\suf_{n_f}(\mathcal{N}(x_{1:t-1}))$
    \STATE $g^{(i)} \leftarrow
      \mathrm{PRNG}\bigl(k_\ast,\, H(c \,\|\, i)\bigr)$;\;\;
      $N_\ast \mathrel{+}= g^{(i)}[x_t]$
  \ENDFOR
\ENDFOR
\STATE $z_\ast \leftarrow
  \bigl(N_\ast - \tfrac{1}{2} T d_\ast\bigr) \big/
  \sqrt{T d_\ast / 4}$ \textbf{for} $\ast \in \{r, f\}$
  \hfill $\triangleright$ Eq.~\eqref{eq:zscore}
\RETURN Intact \textbf{if} $z_r{>}\tau_r \wedge z_f{>}\tau_f$;\;
  Tampered \textbf{if} $z_r{>}\tau_r$;\; \textbf{else} No-Watermark
\end{algorithmic}
\end{algorithm}

\subsection{2D Score Space and Three-State Decision}
Given a text $x = (x_1, \dots, x_T)$, we recover the green-red list $g^{(i)}$ of each round from the hash function and the observed seed, and count the green hits of the rounds assigned to each signal.
Under the null hypothesis of non-watermarked text, each hit is an independent $\mathrm{Bernoulli}(0.5)$ variable, and the count is normalized into a z-score,
\begin{equation}\label{eq:zscore}
  z_\ast = \frac{\sum_{t=1}^{T} \sum_{i : \pi_i = \ast}
    g^{(i)}[x_t] - \tfrac{1}{2} T d_\ast}
  {\sqrt{T d_\ast / 4}},
  \quad \ast \in \{r, f\},
\end{equation}
where $d_r$ and $d_f$ are the numbers of rounds assigned to the robust and the fragile signal, respectively.

The two scores span a 2D score space, and each axis carries one question.
The robust score $z_r$ measures provenance, and the fragile score $z_f$ measures tamper evidence.
A high $z_f$ marks a text as intact on its own, but a low $z_f$ is ambiguous between no-watermark text and text whose fragile seeds have been corrupted by edits, an ambiguity the robust axis resolves.
The decision rule therefore reads $z_f$ only after $z_r$ has attributed the text, partitioning the space into three states,
\begin{equation}
  \mathsf{Det}(x) =
  \begin{cases}
    \text{Intact}   & z_r > \tau_r \text{ and } z_f > \tau_f, \\
    \text{Tampered} & z_r > \tau_r \text{ and } z_f \le \tau_f, \\
    \text{No-Watermark}   & z_r \le \tau_r.
  \end{cases}
\end{equation}
where $\tau_r$ and $\tau_f$ are thresholds set for robust and fragile signal, respectively.
Both intact and tampered texts are attributed to the watermarked LLM, while only intact text is free of tamper evidence.
The ideal separation is as shown in the right panel of Figure~\ref{fig:overview}: 
tampered text scores high on the robust axis and low on the fragile axis, intact text scores high on both, and no-watermark text on neither.

The three-state decision rests on the separability of the three types of text in the 2D space.
When the fragile window is as short as the robust one, the two signals respond identically to edits, and the two scores carry the same information.
The score plane then degenerates into one dimension, and no boundary separates Tampered from Intact.
A fragile window far longer than the robust one is therefore necessary.

Algorithm~\ref{alg:embed} summarizes the embedding of Cocktail, and Algorithm~\ref{alg:detect} summarizes the detection.
The two signals run through one shared pipeline and differ in the window, the key, and the assigned rounds alone.

\section{Experiments}

\subsection{Setup}
We generate with Llama-3.2-1B~\cite{llama3} and Gemma-3-4B~\cite{gemma} on the realnewslike subset of C4~\cite{c4} and on LFQA~\cite{dipper,eli5}, producing 500 texts per model-dataset pair for each scheme.
We compare against four inference-time watermarks.
KGW~\cite{kgw} seeds a balanced green-red split on the previous token and biases green logits.
Unigram~\cite{unigram} fixes one global balanced split.
SynthID~\cite{synthid} seeds on the previous and self token and embeds an unbiased signal through tournament sampling.
SIR~\cite{sir} seeds on a semantic embedding of the prefix and adds a KGW-style bias.
Cocktail uses $h_r=1$, a full-prefix fragile window, $d=30$ rounds, and round ratios $d_r{:}d_f \in \{1{:}1,\, 2{:}1,\, 4{:}1\}$.
Ablation studies generate with Llama-3.2-1B on C4.

\begin{figure}[t]
  \centering
  \includegraphics[width=0.6\columnwidth]{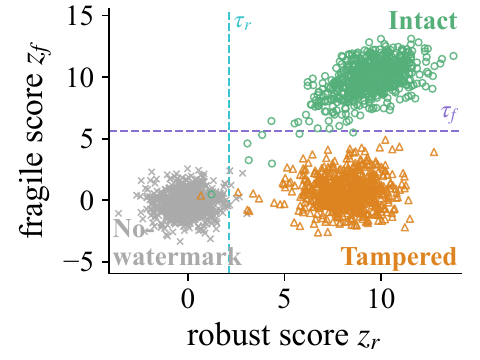}
  \caption{
  Cocktail's score plane $(\zr,\zf)$ on real generations (Llama-3.2-1B, C4). The thresholds $\tau_r,\tau_f$ are each calibrated at a 1\% tail, and the three states read off directly.
  }
  \label{fig:scores}
\end{figure}

\begin{table*}[t]
\centering
\caption{Two-task evaluation on Llama-3.2-1B and Gemma-3-4B.
Attribution reports TPR@1\%FPR for watermarked versus no-watermark text, tamper evidence for tampered versus intact text.
Overall averages the five tasks.
Best in bold, second underlined.
}
\label{tab:main}
\footnotesize
\setlength{\tabcolsep}{5.9pt}
\begin{tabular}{c l ccc cc c c}
\toprule
\multirow{2}{*}{Dataset} & \multirow{2}{*}{Method} & \multicolumn{3}{c}{Attribution $\uparrow$} & \multicolumn{2}{c}{Tamper Evidence $\uparrow$} & Overall & PPL \\
\cmidrule(lr){3-5}\cmidrule(lr){6-7}
 & & No Attack & Paraphrase & RT-Translation & Sentiment-Flip & Token-Substitution & $\uparrow$ & $\downarrow$ \\
\midrule
\multirow{7}{*}{C4 (Llama)}
 & Unigram          & 99.8  & \textbf{98.2} & 78.0 & \phantom{0}3.3 & \phantom{0}0.0 & 55.9 & \phantom{0}9.7 \\
 & KGW              & 100.0 & 86.0 & 76.4 & \phantom{0}5.2 & \phantom{0}6.2 & 54.8 & \phantom{0}8.9 \\
 & SynthID          & 100.0 & 91.2 & 93.8 & \phantom{0}7.5 & \phantom{0}9.8 & 60.5 & 10.6 \\
 & SIR              & 99.0  & 87.4 & 70.0 & \phantom{0}0.0 & \phantom{0}0.0 & 51.3 & \phantom{0}8.8 \\
\rowcolor[gray]{0.93}\cellcolor{white}& Cocktail $1{:}1$ & 99.9  & 82.0 & 93.2 & \textbf{99.4} & \textbf{99.0} & \underline{94.7} & \phantom{0}9.3 \\
\rowcolor[gray]{0.93}\cellcolor{white}& Cocktail $2{:}1$ & 99.9  & 88.4 & \underline{96.2} & \underline{98.3} & \underline{90.5} & \underline{94.7} & \phantom{0}8.8 \\
\rowcolor[gray]{0.93}\cellcolor{white}& Cocktail $4{:}1$ & 99.8  & \underline{92.4} & \textbf{97.4} & 97.2 & 89.5 & \textbf{95.3} & \phantom{0}9.0 \\
\cmidrule(lr){1-9}
\multirow{7}{*}{LFQA (Llama)}
 & Unigram          & 100.0 & \textbf{90.6} & \textbf{94.2} & 17.2 & \phantom{0}6.0 & 61.6 & \phantom{0}7.1 \\
 & KGW              & 100.0 & 78.2 & 75.8 & 22.2 & 12.2 & 57.7 & \phantom{0}7.9 \\
 & SynthID          & 100.0 & 84.8 & \underline{93.8} & 13.0 & \phantom{0}8.8 & 60.1 & \phantom{0}9.3 \\
 & SIR              & 100.0 & \underline{90.1} & 83.3 & 10.7 & \phantom{0}0.0 & 56.8 & \phantom{0}7.7 \\
\rowcolor[gray]{0.93}\cellcolor{white}& Cocktail $1{:}1$ & 100.0 & 80.5 & 84.0 & \textbf{100.0} & \textbf{99.7} & 92.8 & \phantom{0}8.0 \\
\rowcolor[gray]{0.93}\cellcolor{white}& Cocktail $2{:}1$ & 100.0 & 85.7 & 86.3 & \underline{99.4} & \underline{94.3} & \underline{93.1} & \phantom{0}7.9 \\
\rowcolor[gray]{0.93}\cellcolor{white}& Cocktail $4{:}1$ & 100.0 & 89.4 & 88.7 & 98.1 & 91.6 & \textbf{93.6} & \phantom{0}7.5 \\
\midrule
\multirow{7}{*}{C4 (Gemma)}
 & Unigram          & 100.0 & \textbf{95.2} & \textbf{96.2} & \phantom{0}1.8 & \phantom{0}2.7 & 59.2 & \phantom{0}9.0 \\
 & KGW              & 100.0 & 86.0 & 90.6 & \phantom{0}2.4 & \phantom{0}7.7 & 57.3 & \phantom{0}8.6 \\
 & SynthID          & 100.0 & \underline{90.8} & \underline{95.8} & \phantom{0}4.1 & 11.9 & 60.5 & \phantom{0}8.5 \\
 & SIR              & 99.5  & 85.0 & 69.9 & \phantom{0}0.0 & \phantom{0}0.0 & 50.9 & \phantom{0}8.1 \\
\rowcolor[gray]{0.93}\cellcolor{white}& Cocktail $1{:}1$ & 100.0 & 80.8 & 86.4 & \textbf{100.0} & \textbf{99.8} & 93.4 & \phantom{0}8.2 \\
\rowcolor[gray]{0.93}\cellcolor{white}& Cocktail $2{:}1$ & 100.0 & \underline{90.8} & 92.6 & \underline{98.6} & \underline{99.7} & \textbf{96.3} & \phantom{0}8.0 \\
\rowcolor[gray]{0.93}\cellcolor{white}& Cocktail $4{:}1$ & 100.0 & 88.2 & 94.1 & 98.3 & 99.0 & \underline{95.9} & \phantom{0}7.8 \\
\cmidrule(lr){1-9}
\multirow{7}{*}{LFQA (Gemma)}
 & Unigram          & 100.0 & \textbf{90.4} & \textbf{97.0} & \phantom{0}8.0 & \phantom{0}3.5 & 59.8 & 10.2 \\
 & KGW              & 100.0 & 84.4 & 90.2 & 23.1 & 14.8 & 62.5 & \phantom{0}9.9 \\
 & SynthID          & 100.0 & \underline{86.4} & \underline{94.9} & \phantom{0}6.0 & \phantom{0}3.3 & 58.1 & \phantom{0}9.7 \\
 & SIR              & 99.6  & 79.8 & 71.4 & \phantom{0}2.7 & \phantom{0}0.0 & 50.7 & \phantom{0}9.2 \\
\rowcolor[gray]{0.93}\cellcolor{white}& Cocktail $1{:}1$ & 99.7  & 78.9 & 84.6 & \textbf{98.9} & \textbf{98.1} & 92.0 & \phantom{0}8.4 \\
\rowcolor[gray]{0.93}\cellcolor{white}& Cocktail $2{:}1$ & 100.0 & 80.6 & 87.2 & \underline{97.4} & \underline{96.5} & \underline{92.3} & \phantom{0}9.3 \\
\rowcolor[gray]{0.93}\cellcolor{white}& Cocktail $4{:}1$ & 100.0 & 83.1 & 90.3 & 95.7 & 94.2 & \textbf{92.7} & \phantom{0}8.1 \\
\bottomrule
\end{tabular}
\end{table*}

\subsection{Evaluation Protocol}
The two questions of the problem setup induce two discrimination tasks:
(1) \emph{attribution} separates watermarked text, possibly edited, from no-watermark text; 
(2) \emph{tamper evidence} separates tampered watermarked text from intact watermarked text.
The baselines take both tests with their single z-score, and Cocktail takes them with $\zr$ for attribution and $\zf$ for tamper evidence, reading $\zf$ only after $\zr$ has attributed the text.
For each task we report the true positive rate at 1\% false positive rate (TPR@1\%FPR) over the first 300 tokens, thresholded at the 1\% tail of the negative class.
Figure~\ref{fig:scores} shows the calibrated thresholds on the score plane.

The TPR@1\%FPR of each task is the complement of the attacker's success rate.
Attribution faces paraphrasing with Dipper~\cite{dipper} and round-trip translation with opus-mt~\cite{opus-mt}.
Tamper evidence faces random token substitution and a sentiment flip, where gpt-oss:20b~\cite{gpt-oss} flips the sentiment under a word budget and a classifier~\cite{twitter-roberta1} verifies it.
Perplexity (PPL) under a Mistral oracle~\cite{mistral7b} measures generation quality.

\subsection{Attribution and Tamper Evidence}
\paragraph{Cocktail fulfills both goals at no quality cost.}
On the attribution task in Table~\ref{tab:main}, Cocktail attributes unattacked text at 99.7--100.0\%, and under paraphrasing and round-trip translation its $4{:}1$ variant stays within a few points of the strongest single-signal baseline while frequently leading outright.
On the tamper-evidence task, every Cocktail variant flags 89.5--100.0\% of tampered texts at a 1\% false-alarm rate, whereas no baseline exceeds 23.1\%.
Only Cocktail demonstrates over 92\% recall on both goals.
Cocktail's perplexity stays inside the baseline range, confirming that unbiased reweighting preserves generation quality.

\paragraph{The more robust a single signal, the weaker its tamper evidence.}
The baselines order themselves by how insensitive their seeds are to edits.
SIR seeds on a semantic embedding built to survive token-level edits, and delivers 0.0\% tamper evidence in half the settings.
Unigram's green list is fixed and context-independent by construction, and it stays below 17.2\%.
KGW is seeded on the preceding token and peaks at 23.1\%.
SynthID is the most direct control. It uses tournament reweighting but embeds the robust signal only, making it Cocktail with the fragile window degenerated to the robust one, and its tamper evidence does not exceed 13.0\%.

\paragraph{The round ratio dials the trade-off between the two goals.}
Moving from $1{:}1$ to $4{:}1$ on C4/Llama raises paraphrase attribution from 82.0 to 92.4 while tamper evidence recedes from 99.4 to 97.2 under sentiment flip.
The same monotone shift appears in every setting.
The dial moves within a regime where both goals remain far above every baseline: Cocktail's lowest tamper-evidence score of 89.5 still exceeds the best baseline of 23.1 by over 66 points.

\subsection{Ablating the Design Choices}

\paragraph{Tamper evidence requires normalization at seeding time.}
Tamper evidence is defined over the content the reader perceives.
Table~\ref{tab:homoglyph} compares three placements of the normalizer on the same Cocktail watermark under two edits invisible to readers.
Homoglyph substitution~\cite{charperturb} swaps 30\% of eligible characters for lookalike codepoints, e.g., the Latin a (U+0061) for its Cyrillic twin (U+0430).
The None placement seeds and detects the signals on the decoded text.
The Detection placement seeds on the decoded text but detects on the normalized text.
The Seeding placement seeds and detects on the normalized text.
Without normalization, homoglyphs cut attribution to 33.6\% and break the fragile signal on every text.
The attributed texts are flagged as tampered.
We report this flagged fraction as the false alarm on edited texts.
Normalization at detection recovers the attribution but replaces part of the model's original output.
The fragile verification then fails on the replaced intact texts, whose $\zf$ falls to the noise level, and tamper recall under 5\% token substitution collapses from 99.8\% to 43.5\%.
Normalizing before seeding gives both sides the same canonical form, and the edited texts enter detection as their intact counterparts.
At a 1\% false-alarm rate, tamper recall stays at 98.3\%.

\begin{table}[!t]
\centering
\caption{Attribution and tamper evidence under homoglyph substitution across three normalizer placements.}
\label{tab:homoglyph}
\footnotesize
\setlength{\tabcolsep}{4pt}
\begin{tabular}{l c c c}
\toprule
& \multicolumn{3}{c}{Normalized at} \\
\cmidrule(lr){2-4}
& None & Detection & Seeding \\
\midrule
Attribution $\uparrow$    & 33.6 & 99.6 & 100.0 \\
False alarm (edited) $\downarrow$ & 33.6 & 38.3 & \phantom{0}1.0 \\
Tamper recall $\uparrow$  & 99.8 & 43.5 & 98.3 \\
\bottomrule
\end{tabular}
\end{table}

\paragraph{Co-embedding outperforms one signal per token.}
To achieve the two goals, a natural alternative assigns each \emph{token} wholly to one signal, keeping each signal's expected total budget unchanged.
Yet this assignment fails once the detector loses track of which signal each token carries.
We therefore grant the variant its best detector and score every token under both signal keys.
For each signal, half of the scored positions carry only the other signal's noise.
The intact $\zr$ drops from 9.4 to 6.1, and matching the co-embedded evidence takes roughly twice the text.
Attribution at $T{=}50$ falls from 98.3\% to 86.0\%, and round-trip attribution at the standard length falls from 89.8\% to 60.2\%.
Hence, co-embedding is inherent to achieving the two goals jointly.

\begin{figure}[t]
  \centering
  \includegraphics[width=\columnwidth]{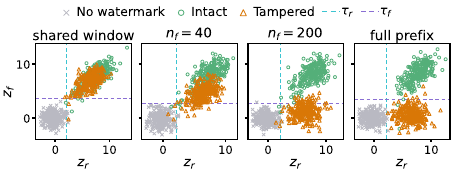}
  \caption{Score planes under four fragile windows. Intact and tampered text separate on $\zf$ only as the window grows.}
  \label{fig:windowsep}
\end{figure}

\paragraph{Tamper evidence requires a long fragile window and attribution first.}
Figure~\ref{fig:windowsep} sweeps the fragile window length on the score plane.
When the fragile window shrinks to the robust one, the fragile signal survives the same edits as the robust one and Cocktail degenerates into SynthID.
Intact and tampered text collapse onto one diagonal cloud.
Lengthening the character window peels the tampered cloud off the intact one.
Tamper evidence reaches 7.0\% at $n_f{=}40$ and 88.0\% at $n_f{=}200$.
Identifying tampering requires the text to be attributed to the watermarked LLM first.
Tampered and no-watermark text both sit near zero on the $\zf$ axis, and only the tampered text keeps a high $\zr$.
Reading $\zf$ only after $\zr$ has attributed the text therefore separates the two.

\subsection{Comparison with Signature-Based Schemes}
Bileve~\cite{bileve}, the closest design to Cocktail, pairs a coarse statistical signal for provenance with a digital signature for integrity.
Verification requires exact bit recovery, which exacts a price on what is signed and how the bits are carried.
We run the official implementation with default parameters on Llama-3.2-1B and C4 to measure the price.
\paragraph{Signing does not certify the delivered text.}
Bileve signs token IDs, and re-tokenization of the delivered text fails verification on 91.1\% of unmodified texts.
Verified on the stored token IDs instead, the same pipeline passes every text, attributing the failures to re-tokenization.
Signature verification leaves no threshold to tune.
The integrity channel operates at an effective false-alarm rate of 91.1\%, against Cocktail's calibrated 1\%.
The signed message also stops at the first 44 tokens, and the following 512 tokens merely carry the signature bits without enjoying its certification.
Cocktail instead seeds the fragile signal on the normalized prefix, anchoring tamper evidence to the delivered text itself.

\paragraph{Carrying exact bits costs quality and compute.}
To carry each bit intact, Bileve takes the top-ranked token whose hash matches it, overriding sampling even at low-entropy positions.
Perplexity rises to 62.2, six times Cocktail's 10.3 under the identical measurement protocol.
Recovering the bits is equally expensive.
The detector realigns the text against the key sequence through permutation tests, whereas Cocktail scores in closed form.
The permutation test reports $p=(c+1)/(n_{\mathrm{runs}}+1)$, and the default 20 permutations bound $p\ge 1/21$, above our 1\% operating point. Scoring one text takes 144 s even with our accelerated alignment kernel, and reaching 1\% needs at least 99 permutations.

\section{Discussion and Conclusion}
A signal that survives edits cannot also expose them, leaving single-signal watermarks open to piggyback spoofing.
Cocktail escapes this conflict by co-embedding a robust and a fragile signal into every token through unbiased reweighting, both seeded on the text the reader receives, and read jointly as a three-state decision.
Across two models and two datasets, it flags 89.5 to 100\% of tampered texts at a 1\% false-alarm rate without sacrificing attribution or perplexity.
Edits confined to the final tokens corrupt only the tail of the fragile seeds, and reading $\zf$ over trailing segments is a natural extension we leave to future work.

Besides piggyback spoofing, a threat to watermarks is stealing, which forges the signal by inferring the green list through queries~\cite{watermark_stealing}.
The cost of this inference grows with the seeding window~\cite{unforgeable_wm,dual_robust_spoof}.
Cocktail's short robust window shares this exposure with existing robust schemes, whereas the fragile signal seeds on the full normalized prefix and remains hard to forge.
The signal that supplies tamper evidence is thus the harder one to steal, leaving forgeries flagged as Tampered, and the recipe of two complementary signals may extend to other generative modalities where provenance and tamper evidence both matter.

\appendices

\section{Unbiasedness and Entropy Consumption of Tournament Reweighting}

Cocktail introduces no new theory.
The unbiasedness and budget claims of the main paper rest on theorems proved in the Supplementary Information of SynthID~\cite{synthid}.
To keep this appendix self-contained, the statements below reproduce those theorems and their supporting definitions with the original numbering.
Throughout, ``Methods Algorithm~2'' and ``Methods Algorithm~3'' refer to the algorithms of those names in SynthID~\cite{synthid}, and $\mathcal{V}$ denotes the vocabulary.

\subsection{Notation restated from SynthID}

The following definitions appear in Supplementary Appendices~E, G.1, G.2, and~H.1 of SynthID~\cite{synthid}.
Here $\mathcal{R}$ is the space of random seeds, $\Delta \mathcal{V}$ is the probability simplex over $\mathcal{V}$, and $g_\ell(x, r)$ is the layer-$\ell$ pseudorandom $g$-value of token $x$ under seed $r$ (Methods Definition~4 of SynthID~\cite{synthid}), whose marginal distribution is $f_g$.

\medskip
\noindent\textbf{Definition 9} (Watermarked distribution).
\emph{Given a probability distribution $p$ over $\mathcal{V}$, a random seed $r \in \mathcal{R}$, a number of samples $N \ge 2$, a $g$-value distribution $f_g$, and a number of layers $m \ge 1$, the \textnormal{watermarked distribution} $p_{wm}(\cdot \mid p, r, f_g, N, m)$ is the probability distribution of the winner of Methods Algorithm~2:}
\begin{equation}
\begin{split}
&p_{wm}(x_t \mid p, r, f_g, N, m) \\
&\qquad\qquad= \mathbb{P}\left[\mathrm{Alg2}(p, r, f_g, N, m) \text{ returns } x_t\right].
\end{split}
\end{equation}

\medskip
\noindent\textbf{Definition 16} (Single-token non-distortionary sampling algorithm).
\emph{A sampling algorithm $\mathcal{S} : \Delta \mathcal{V} \times \mathcal{R} \to \mathcal{V}$ is \textnormal{(single-token) non-distortionary} if for any probability distribution $p \in \Delta \mathcal{V}$ and token $x \in \mathcal{V}$:}
\begin{equation}
\mathbb{E}_{r \sim \mathrm{Unif}(\mathcal{R})}\left[\mathbb{P}\left(\mathcal{S}(p, r) = x\right)\right] = p(x).
\end{equation}
\emph{If $\mathcal{S}$ is not non-distortionary, we call it \textnormal{distortionary}.}

\medskip
\noindent
In Definition~20 the notation
$\mathbb{P}_{wm}\bigl(\mathbf{y}^i \mid \mathbf{x}^i, k;\, \cdot\,\bigr)$ denotes the probability that the watermarking scheme with key $k$ generates response $\mathbf{y}^i$ to prompt $\mathbf{x}^i$ conditioned on the earlier prompt and response pairs $(\mathbf{x}^1, \mathbf{y}^1), \ldots, (\mathbf{x}^{i-1}, \mathbf{y}^{i-1})$, and $\mathcal{V}^*$ is the set of all finite sequences in $\mathcal{V}$.

\medskip
\noindent\textbf{Definition 20} ($K$-sequence non-distortionary watermarking scheme).
\emph{A watermarking scheme $\mathbb{P}_{wm}$ is \textnormal{$K$-sequence non-distortionary} for some $K \ge 1$ if, for any sequence of $K$ prompts $\mathbf{x}^1, \ldots, \mathbf{x}^K \in \mathcal{V}^*$ and sequence of $K$ responses $\mathbf{y}^1, \ldots, \mathbf{y}^K \in \mathcal{V}^*$:}
\begin{equation}
\begin{split}
&\mathbb{E}_{k \sim \mathrm{Unif}(\mathcal{R})}\Bigl[\prod_{i=1}^{K} \mathbb{P}_{wm}\bigl(\mathbf{y}^i \mid \mathbf{x}^i, k;
(\mathbf{x}^1, \mathbf{y}^1), \ldots, (\mathbf{x}^{i-1}, \mathbf{y}^{i-1})\bigr)\Bigr] \\
&\qquad\qquad= \prod_{i=1}^{K} p_{LM}(\mathbf{y}^i \mid \mathbf{x}^i).
\end{split}
\end{equation}
\medskip
\noindent\textbf{Definition 22} (Collision probability).
\emph{Given a probability distribution $p$, the \textnormal{collision probability} $C_p$ of $p$ is the probability that two samples drawn i.i.d.\ from $p$ are the same. If $p = (p_i)_{i=1}^{N}$ is discrete, the collision probability equals $\sum_{i=1}^{N} p_i^2$.}

\medskip
\noindent
The SynthID supplement adds that collision probability is related to collision entropy, sometimes called R\'enyi entropy, $H_2(p) = -\log \sum_{i=1}^{N} p_i^2$.

\medskip
\noindent\textbf{Definition 23} (Higher-order collision probabilities).
\emph{Given a probability distribution $p$ and integers $N, j \ge 1$, let $C_p^{N,j}$ denote the probability that $N$ samples drawn i.i.d.\ from $p$ have exactly $j$ unique values. Note that $C_p^{2,1}$ is the collision probability of $p$. In general, we refer to $C_p^{N,j}$ as the \textnormal{higher-order collision probabilities} of $p$.}

\medskip
\noindent\textbf{Definition 24} (Watermarked $g$-value distribution).
\emph{Given a probability distribution $p$, a $g$-value distribution $f_g$, and number of samples $N \ge 2$, let $F_{gw}$ denote the cumulative density function of the $g$-value of a token sampled from the single-layer watermarked distribution $p_{wm}(\cdot \mid p, r, f_g, N, 1)$ (Definition~9), in expectation over the random seed $r$:}
\begin{equation}
F_{gw}(z) := \mathbb{P}_{r \sim \mathrm{Unif}(\mathcal{R}),\, x \sim p_{wm}(\cdot \mid p, r, f_g, N, 1)}\left[g_1(x, r) \le z\right].
\end{equation}
\emph{Let $f_{gw}$ denote the probability density/mass function corresponding to $F_{gw}$. We refer to $f_{gw}$ as the \textnormal{watermarked $g$-value distribution}.}

\subsection{Unbiasedness of multi-round reweighting}

Theorems 17 and 18 below appear in Supplementary Appendix~G.1 of SynthID~\cite{synthid} and Theorem 21 in its Appendix~G.2.

\medskip
\noindent\textbf{Theorem 17} (Single-layer two-sample Tournament sampling is non-distortionary).
\emph{For any probability distribution $p$ over $\mathcal{V}$, $g$-value distribution $f_g$, and token $x_t \in \mathcal{V}$:}
\begin{equation}
\mathbb{E}_{r_t \sim \mathrm{Unif}(\mathcal{R})}\left[p_{wm}(x_t \mid p, r_t, f_g, 2, 1)\right] = p(x_t).
\end{equation}

\medskip
\noindent\textbf{Theorem 18} (Multi-layer two-sample Tournament sampling is non-distortionary).
\emph{For any probability distribution $p$ over $\mathcal{V}$, $g$-value distribution $f_g$, number of layers $m \ge 1$, and token $x_t \in \mathcal{V}$:}
\begin{equation}
\mathbb{E}_{r_t \sim \mathrm{Unif}(\mathcal{R})}\left[p_{wm}(x_t \mid p, r_t, f_g, 2, m)\right] = p(x_t).
\end{equation}

\medskip
\noindent\textbf{Theorem 21} ($K$-sequence repeated context masking + non-distortionary sampling algorithm $\to$ $K$-sequence non-distortionary watermarking scheme).
\emph{Let $\mathcal{S}$ be a non-distortionary sampling algorithm (Def 16). For any $K \ge 1$, let $\mathbb{P}_{wm}$ denote the watermarking scheme that applies $\mathcal{S}$ with sliding window random seed generation and $K$-sequence repeated context masking (Methods Algorithm 3). Then $\mathbb{P}_{wm}$ is $K$-sequence non-distortionary.}

\medskip
\noindent
Our vectorized tournament, main-paper Eq.~3, is the closed form of the same two-sample tournament with a $\mathrm{Bernoulli}(0.5)$ list, and our rounds satisfy the independence premise of Theorem 18: round $i$ draws its list as $g^{(i)}=\mathrm{PRNG}(k_{\pi_i}, H(c \,\|\, i))$, and distinct round indices and keys yield independent lists under the pseudorandomness of $H$.
Our embedding and detector skip a token whenever the context hash of either window repeats within the generation, which is the repeated context masking that Theorem 21 pairs with a non-distortionary sampler for the sequence-level guarantee (the $K=1$ instantiation, i.e., single-sequence non-distortion), a requirement also standard in unbiased watermarking~\cite{unbiased}.
Unbiasedness therefore follows.

\subsection{Entropy consumption across rounds}

The three theorems below appear in Supplementary Appendices~H.3 and~H.4 of SynthID~\cite{synthid}, where the LLM distribution is written $p_{LM}$.

\medskip
\noindent\textbf{Theorem 29} ($g$-value bias increases with $N$, single-layer tournament).
\emph{Given a probability distribution $p_{LM}$ and $g$-value distribution $f_g$, let $F_{gw}^{N}$ be the c.d.f.\ of the watermarked $g$-value distribution for a single-layer tournament with $N$ samples. Let $F_{gw}^{N+1}$ be the same for a single-layer tournament with $N+1$ samples. Then for all $z$:}
\begin{equation}
F_{gw}^{N+1}(z) \le F_{gw}^{N}(z).
\end{equation}
\emph{When $0 < F_{gw}^{N}(z) < 1$, equality holds iff $p_{LM}$ is one-hot.}

\medskip
\noindent\textbf{Theorem 31} (Expected collision probability for single-layer tournament, two samples).
\emph{Given a probability distribution $p_{LM}$, random seed $r \in \mathcal{R}$ and $g$-value distribution $f_g$, let $C^{2,1}_{p_{wm}}$ denote the collision probability of the watermarked distribution $p_{wm}(\cdot \mid p_{LM}, r, f_g, 2, 1)$ for a $N=2$ sample single-layer tournament. In expectation over the random seed $r$, the collision probability is:}
\begin{equation}
\begin{split}
\mathbb{E}_{r \sim \mathrm{Unif}(\mathcal{R})}\left[C^{2,1}_{p_{wm}}\right] &= \left[\tfrac{4}{3} - \tfrac{1}{3} C^{3,1}_{f_g}\right] C^{2,1}_{p_{LM}} \\
&+ \left[\tfrac{2}{3} + \tfrac{1}{3} C^{3,1}_{f_g} - C^{2,1}_{f_g}\right] \left(C^{2,1}_{p_{LM}}\right)^2 \\
&- \left[\tfrac{2}{3} - \tfrac{2}{3} C^{3,1}_{f_g}\right] C^{3,1}_{p_{LM}} \\
&- \left[\tfrac{1}{3} + \tfrac{2}{3} C^{3,1}_{f_g} - C^{2,1}_{f_g}\right] C^{4,1}_{p_{LM}}.
\end{split}
\end{equation}
\emph{where $C^{N,j}_{p_{LM}}$ and $C^{N,j}_{f_g}$ are the higher order collision probabilities (Def~23), respectively, of $p_{LM}$ and $f_g$.}

\medskip
\noindent\textbf{Theorem 32} (Single-layer tournament increases the expected collision probability, two samples).
\emph{The expected collision probability of a single-layer tournament with $N=2$ samples is greater than or equal to the LLM collision probability: $\mathbb{E}_{r \sim \mathrm{Unif}(\mathcal{R})}\left[C^{2,1}_{p_{wm}}\right] \ge C^{2,1}_{p_{LM}}$, with equality iff $p_{LM}$ is one-hot.}

\medskip
\noindent
Supplementary Appendix~H.4 of SynthID~\cite{synthid} draws the multi-layer consequence: applied layer after layer, the tournament produces distributions whose expected collision probability keeps rising, each new layer therefore contributes less watermarking strength than the one before, and adding layers can yield diminishing returns.

\medskip
\noindent
Read in our notation, the collision entropy of $\hat p^{(i)}$ decreases monotonically in $i$, and every round spends a share of the remaining budget.
This grounds the round-allocation dial of the main paper: rounds assigned at a ratio $(r{:}1)$ split the shrinking budget between the two signals roughly at that ratio, and the ratio acts as a monotone knob rather than an exact linear control.

\begin{table}[h]
\centering
\caption{Hyperparameters of all methods and evaluations.}
\label{tab:hyperparams}
\footnotesize
\begin{tabular}{l l}
\toprule
Item & Value \\
\midrule
\multicolumn{2}{l}{\emph{Cocktail}} \\
Robust window $h_r$ & 1 preceding token + self token \\
Fragile window $n_f$ & full normalized prefix + self token \\
Rounds $d$ & 30 \\
Round ratios $d_r{:}d_f$ & $1{:}1$, $2{:}1$, $4{:}1$ \\
Fragile seed & HMAC-SHA256 over the prefix \\
Robust seed and layers & SynthID accumulate\_hash \\
\midrule
\multicolumn{2}{l}{\emph{Baselines}} \\
KGW & $\gamma=0.5$, $\delta=2.0$, window 1 \\
Unigram & $\gamma=0.5$, $\delta=2.0$, global list \\
SynthID & $m=30$ layers \\
 & 1 preceding token + self token \\
SIR & compositional-bert-large, $\delta=1.0$ \\
\midrule
\multicolumn{2}{l}{\emph{Attacks}} \\
Paraphrase (Dipper) & lex 20/60 with order 0 \\
 & order 20/60 with lex 0 \\
Round-trip translation & opus-mt en$\to$fr$\to$en \\
Token substitution & $\rho \in \{5\%, 10\%\}$, mean reported \\
Sentiment flip & gpt-oss:20b, word budgets 5--30\% \\
Sentiment judge & twitter-roberta-base-sentiment \\
NLI filter & DeBERTa-xlarge-MNLI \\
Homoglyph substitution & 30\% of eligible characters \\
\midrule
\multicolumn{2}{l}{\emph{Quality}} \\
PPL oracle & Mistral-7B-v0.1, first 200 tokens \\
\bottomrule
\end{tabular}
\end{table}

\section{Implementation Details}

\paragraph{Computing infrastructure.}
All experiments run on a single NVIDIA RTX 4090 Laptop GPU with 16 GB of VRAM, an Intel Core i9-14900HX CPU, and 64 GB of RAM, under Windows 11 Pro.
The implementation uses Python 3.10, PyTorch 2.10 with CUDA 12.6, and transformers 4.57.
Baselines run through the released code: the SynthID tournament reuses the hashing kernel of the official synthid-text release, and SIR loads the transform network and token mapping from its official repository.

\paragraph{Generation and detection configuration.}
Each scheme generates 500 texts of up to 512 tokens per model-dataset pair with pure sampling at temperature 1.0 over the top 100 logits.
C4 prompts are the first 100 words of realnewslike documents, streamed from the validation split shuffled with seed 42 after skipping 5000 documents, and LFQA ships its prompts.
Token sampling is not seeded.
Each run records its configuration in generation\_params.json, detection reads it back, and scoring the released texts is exactly reproducible.
Detectors score the first 300 tokens of the delivered text, and the sentiment-flip evaluation scores 200 because its inputs are truncated to 200 words before rewriting.
Thresholds are calibrated per setting: $\tau_r$ at the 99th percentile of no-watermark $z_r$, and $\tau_f$ at the 1st percentile of intact $z_f$.

\begin{figure}[!h]
\centering
\includegraphics[width=\columnwidth]{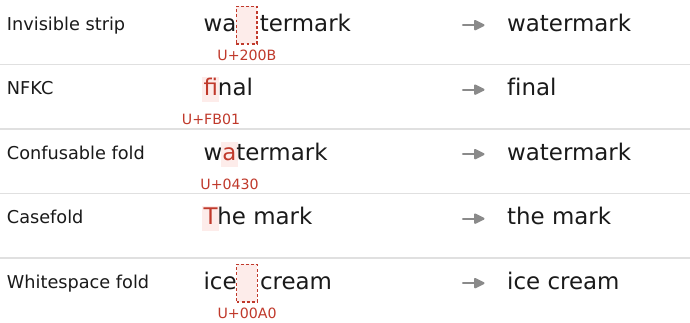}
\caption{Table~\ref{tab:normalizer}'s examples. Red marks the folded codepoint, and dashed boxes mark characters print cannot reveal.}
\label{fig:normalizer}
\end{figure}

\begin{table}[h]
\centering
\caption{Normalizer operations.}
\label{tab:normalizer}
\footnotesize
\setlength{\tabcolsep}{4pt}
\begin{tabular}{l l l}
\toprule
Operation & Cancels & Example \\
\midrule
Invisible strip & zero-width chars & \texttt{a[U+200B]b} $\to$ \texttt{ab} \\
\textsc{nfkc} & compatibility forms & \texttt{[U+FB01]} $\to$ \texttt{fi} \\
Confusable fold & homoglyphs & \texttt{[U+0430]} $\to$ \texttt{a} \\
Casefold & case changes & \texttt{The} $\to$ \texttt{the} \\
Whitespace fold & spacing edits & \texttt{[U+00A0]} $\to$ space \\
\bottomrule
\end{tabular}
\end{table}

\begin{figure*}[t!]
\centering
\footnotesize
{\setlength{\fboxsep}{6pt}%
\fcolorbox{framegray}{white}{%
\begin{minipage}{0.97\textwidth}
\setlength{\fboxsep}{1.4pt}%
\textbf{($\star$) Cocktail-watermarked output of Gemma-3-4B} (\textcolor[HTML]{59B292}{Intact})\hfill detector output: \textbf{Intact}, $z_r{=}11.29>\tau_r$, $z_f{=}9.63>\tau_f$\\[3pt]
\ldots\ And you know in that moment you're wearing your best swimsuit and you have a sunburn, and you are happy.\ \ldots\ Miami has a veritable tiki bar renaissance\ \ldots\ But tiki is not dead: It's just not going to your neighborhood bar.\ \ldots
\par\vspace{6pt}
{\centering \emph{sentiment flip of ($\star$): nine word substitutions convert the statement while preserving attribution}\par}
\vspace{4pt}
\textbf{Sentiment-flip attack} (\textcolor[HTML]{E07A2F}{Tampered})\hfill detector output: \textbf{Tampered}, $z_r{=}10.63>\tau_r$, $z_f{=}{-}0.78<\tau_f$\\[3pt]
\ldots\ And you know in that moment you're wearing your \etok{worst} swimsuit and you have a sunburn, and you are \etok{unhappy.}\ \ldots\ Miami has a \etok{sham} tiki bar \etok{decline}\ \ldots\ But tiki \etok{is dead:} It's just not going to your neighborhood bar.\ \ldots
\par\vspace{6pt}
{\centering \emph{homoglyph substitution of ($\star$): lookalike codepoints replace 81 characters, invisible to the reader}\par}
\vspace{4pt}
\textbf{Homoglyph attack} (\textcolor[HTML]{59B292}{Intact})\hfill detector output: \textbf{Intact}, scores unchanged\\[3pt]
\ldots\ \htok{And} you \htok{know} in \htok{that} \htok{moment} \htok{you're} wearing your \htok{best} \htok{swimsuit} and you \htok{have} \htok{a} sunburn, \htok{and} \htok{you} \htok{are} happy.\ \ldots
\par
\end{minipage}}}
\caption{
    One Gemma-3-4B generation under a sentiment-flip and a homoglyph attack. Orange boxes mark word substitutions. Blue boxes mark homoglyph swaps, printed as the corresponding Latin characters. Seeding on normalized text leaves the scores unchanged, while seeding on raw tokens drops $z_r$ to 1.81.
}
\label{fig:qualitative}
\end{figure*}

\paragraph{Normalizer specification.}
The normalizer strips zero-width and invisible codepoints, applies Unicode \textsc{nfkc}, folds visual confusables to ASCII prototypes via the UTS~39 skeleton and a fixed Cyrillic and Greek to Latin override table, casefolds, and collapses whitespace to single spaces, in that order.
Table~\ref{tab:normalizer} pairs each operation with the surface channel it cancels, and Figure~\ref{fig:normalizer} renders the same examples with their true glyphs.
Each operation is idempotent, and the composition is deterministic across platforms.
An edit that escapes the normalizer changes the normalized text itself, and detection reports it as tampering rather than passing it silently.

\section{Bileve Measurement Protocol}

We run the official Bileve implementation with its default parameters: ECDSA over P-256, message length $d=44$ tokens, carrier length $m=512$ tokens, alignment block $n=80$.
The evaluation covers 500 intact watermarked texts and 500 no-watermark texts on Llama-3.2-1B with C4 prompts.

\paragraph{Signature channel.}
Verified on the delivered text after re-tokenization, the signature passes on 8.9\% of intact texts.
Verified on the stored token IDs, the same pipeline passes on 100\% of intact and 0\% of no-watermark, tampered, and paraphrased texts.
The failures therefore stem from re-tokenization, not from our reproduction.
\paragraph{Detection cost.}
One pass aligns the 433 length-80 windows of the 512-token carrier against the key sequence at 80 candidate offsets, about $3.5\times10^{4}$ Levenshtein alignments.
The permutation p-value at the default $n_{\mathrm{runs}}=20$ runs this pass once with the true key and 20 more times with random keys, and every channel whose signature fails to verify pays the full 21 passes.
We compile the official Cython alignment kernel to native code and replace its per-call 41 MB key-matrix copy with zero-copy views, leaving the alignment count itself as the remaining cost.
The 144 s per text reported in the main paper is measured after these optimizations.

\section{Examples of Cocktail-Watermarked Text}

Figure~\ref{fig:qualitative} shows the delivered text, the edits, and the scores $(z_r, z_f)$ with the three-state decision, for one Gemma-3-4B generation on a C4 prompt, scored over the first 200 tokens.
Nine word edits out of 200, confirmed by the NLI filter as a positive to negative flip, leave the robust score nearly unchanged and collapse the fragile score, the score pattern of piggyback spoofing that the joint rule catches.
The homoglyph attack swaps 81 of the 314 eligible characters for lookalike codepoints and is invisible to the reader.
Seeded on raw tokens, the attacked text scores ($z_r=1.81$, $z_f=-1.23$) and loses attribution.
Seeded on normalized text, the identical attack becomes ineffective.

\bibliographystyle{IEEEtran}
\bibliography{aaai2027}

\end{document}